\documentclass[12pt,twoside]{article}
\usepackage{graphicx}
\usepackage[small,bf,sf,textfont={small,sf,bf}]{caption}
\usepackage{subfig}
\usepackage{fancyheadings}
\usepackage[numbers]{natbib}
\usepackage{algorithm}
\usepackage{algorithmic}
\usepackage{amsmath}
\usepackage{amsfonts}
\usepackage{amsthm}
\usepackage{float}
\usepackage{mathrsfs}
\usepackage{pslatex}
\usepackage{color}
\usepackage{enumitem}
\usepackage{listings}
\usepackage{fancyvrb}
\usepackage{fancyheadings}

\newcommand\rev[1]{{\color{black}#1}}

\begin{document}

\pagestyle{fancy}

\rfoot[]{}
\cfoot[]{}
\lfoot[]{}

\setlength{\columnseprule}{0pt}

\newcommand{\trp}{^{\scriptsize \text{T}}}
\newcommand{\itp}{^{\scriptsize -\text{T}}}
\newcommand{\sqrtp}{^{\scriptsize \text{T/2}}}
\newcommand{\isqrtp}{^{\scriptsize -\text{T/2}}}
\newcommand{\inv}{^{\scriptsize -1}}
\setlist[enumerate]{itemsep=0mm}
\newtheorem{example}{Example}

\definecolor{verbgray}{gray}{0.9}

\lstnewenvironment{mvb}{%
      \lstset{backgroundcolor=\color{verbgray},
        frame=single, framerule=0pt, basicstyle=\scriptsize\ttfamily,
              columns=fullflexible, keepspaces=true}}{}

\thispagestyle{empty}

{\noindent \sffamily \Large
\textbf{Load Balancing using Hilbert Space-filling Curves for Parallel Reservoir Simulations}
}

{\noindent \sffamily Hui Liu, Kun Wang, Bo Yang, Min Yang, Ruijian He, Lihua Shen,
He Zhong, Zhangxin Chen\\ University of Calgary}
\bigskip

\noindent \hrulefill

\section*{Abstract}
New reservoir simulators designed for parallel computers enable us to overcome performance
limitations of personal computers and to simulate large-scale
reservoir models. With development of parallel reservoir simulators, more
complex physics and detailed models can be studied. The key to design efficient parallel reservoir
simulators is not to improve the performance of individual CPUs drastically but to utilize the aggregation of
computing power of all requested nodes through high speed networks \cite{larry-f}. An ideal scenario is
that when the number of MPI (Message Passing Interface) \rev{processes} is doubled, 
the running time of parallel reservoir
simulators is reduced by half.

The goal of load balancing (grid partitioning) is to minimize overall \rev{computations and communications},
and to make
sure that all \rev{processors} have a similar workload \cite{sfc-tr-03}.
Geometric methods divide a grid by using a location of \rev{ a cell} \cite{zoltan-ug}
while topological methods work with connectivity of cells, which is generally described as a graph \cite{parmetis}.
This paper introduces a Hilbert
space-filling curve method. A space-filling curve is a continuous curve and defines a map
between a one-dimensional space and a multi-dimensional space \cite{Hans}. A Hilbert space-filling curve is
one special space-filling curve discovered by Hilbert and has many useful characteristics \cite{Hans, butz},
such as good locality, which means that two objects that are close to each other in a multi-dimensional
space are also close to each other in a one dimensional space. This property can model communications
in grid-based parallel applications. The idea of the Hilbert space-filling curve method is to map a
computational domain into a one-dimensional space, partition the one-dimensional space to
certain intervals, and assign all cells in a same interval to a MPI \rev{process}.
To implement a load balancing method, \rev{a mapping kernel is required to convert}
high-dimensional coordinates to a scalar value and an efficient one-dimensional partitioning
module that divides a one-dimensional space and makes sure that all intervals have a similar workload.

\rev{The Hilbert space-filling curve method is compared with ParMETIS, a famous graph partitioning package.
The results show that our Hilbert space-filling curve method has good partition quality.
It has been applied to grids with billions of cells, and linear scalability has been obtained on
IBM Blue Gene/Q.
}

\section{Introduction}

Reservoir simulators are powerful tools in the oil and gas industry, \rev{which have been applied} to
validate production designs, to predict future performance of wells and reservoirs, and
to enhance oil recovery.
The demand for modeling capability for petroleum reservoirs raises with a dramatic increase in
computational \rev{power}, as the physical and chemical processes require more accurate assessment
and the geological models become more complex \cite{larry-f}.
As a consequence, it may take days or even weeks to complete one simulation run by personal computers.
Effective numerical methods and parallel reservoir simulators have been studied
to solve this issue.

Reservoir simulations have been popular research topics since the 1950s.
\rev{Since then, various models,} such as the black oil, compositional and thermal models, have been proposed,
and their numerical methods have been studied.
\rev{
    Many commercial reservoir simulators have been developed, most of which were designed for
personal computers.
}
However, the performance of personal computers is limited by their CPUs and memory size.
Parallel computers, which are a group of \rev{computers (nodes) connected by high
speed networks, such as InfiniBand},
have been employed to accelerate reservoir simulations.
\rev{MPI (Message Passing Interface)} is commonly accepted as a communication environment.
\rev{It is well-known that the scalability of a parallel application is determined by
communication and the nature of the application \cite{amdahl}, such as the proportion of execution time
that can be parallelized in the application. The keys are to identify the parallelizable portion
and to reduce communication among these nodes.}
For grid-based methods, such as the finite element,
finite volume and finite difference methods, the communication pattern of a parallel application
can be determined by a given grid and its numerical method,
and \rev{it can} be described by a graph. Many graph-based grid partitioning algorithms have been proposed,
including spectral methods \cite{nc2,nc9}, multilevel methods \cite{nc10,nc12},
\rev{
    diffusive methods \cite{nc13,nc14,nc15},
    and refinement-tree methods \cite{HSFC,mitchell-k,mitchell-hp},
    whose partitioning quality is excellent.
    These algorithms have been implemented
    and are available online. METIS and ParMETIS \cite{metis,parmetis}, Scotch and PTScotch \cite{scotch,ptscotch},
    Chaco \cite{chaco} and Jostle \cite{jostle} are famous graph partitioning packages.
    However, one disadvantage of the graph methods is that they are complex, and they are hard to implement
    if a load balancing module must be implemented. As an alternative, geometric methods have been studied,
    which assume that objects have a higher
    chance to communicate with nearby objects \cite{sfc-tr-03}. Their {partitioning} quality is also good and easier to implement,
    and they are good alternatives for graph methods.
    Therefore, geometric methods are also popular in parallel computing.
}
\rev{Many geometric} algorithms have been developed for them, including space-filling curves (Hilbert space-filling curves \cite{HSFC} and
\rev{Morton space-filling curves \cite{sfc-tr-03}), recursive bisection \cite{nc1,nc2,nc3},
and Zoltan \cite{zoltan-dev,zoltan-ug}.
}

\rev{
    This paper introduces our work on applying the Hilbert space-filling curve method to large-scale reservoir simulations
    and studying its partitioning quality and its scalability of reservoir simulations.
    The finite difference (volume) methods are applied and a structured grid is
    attached to each reservoir. The communication pattern of parallel reservoir simulations is determined by the partitioning
    of grids and numerical methods. In Section \S \ref{sec-part}, grid partitioning and communication
    modeling are introduced. In Section \S \ref{sec-sfc}, space-filling curves and Hilbert
    space-filling curve orders are presented. In Section \S \ref{sec-sfcp}, the Hilbert space-filling curve
    partitioning method and partitioning quality are introduced. Section \S \ref{sec-num} shows
    numerical experiments to study partitioning methods.
}

\section{Grid Partitioning}
\label{sec-part}

\rev{Assume} that $N_p$ MPI tasks exist.
Let $\Omega = [x_1, x_2] \times [y_1, y_2] \times [z_1, z_2]$
be the reservoir domain, and the domain has been partitioned into $n_x$, $n_y$ and $n_z$ intervals in the $x$, $y$ and $z$ directions,
respectively.
Let $\mathbb{G}$ be this structured grid, which has $N_g = n_x \times n_y \times n_z$ cells.
\rev{$\mathbb{G}$ is used} to represent the union of all cells,
\begin{equation}
\mathbb{G} = \{B_1, B_2, \cdots, B_{N_g}\},
\end{equation}
where $B_i$ is the $i$th cell of $\mathbb{G}$. The grid $\mathbb{G}$ is distributed \rev{to} $N_p$ MPI
\rev{processes}, and each \rev{process} has
a subset of $\mathbb{G}$. Let $\mathbb{G}_i$ be the sub-grid owned by the $i$th \rev{process}, which
satisfies the following conditions:
\begin{equation}
 \left\{
 \begin{aligned}
& \mathbb{G}_i \neq \emptyset \ (i = 0, \cdots, N_p - 1), \\
& \mathbb{G}_i \cap \mathbb{G}_j = \emptyset \ (i \neq j), \\
& \cup \mathbb{G}_i = \mathbb{G} \ (i = 0, \cdots, N_p - 1).
\end{aligned}
 \right.
\end{equation}
Any cell $B_i$ belongs to some sub-grid, whose
neighboring cells may belong to different sub-grids.

\subsection{Communication Modeling}

Information from a neighbor is always required for the finite difference, finite volume
and finite element methods. For example, in the finite difference method,
equations (\ref{fo-forward}), (\ref{ba-forward}) and (\ref{ca-forward}) are
used to calculate the first-order derivative, and a value from another cell (point) is always required.
Equation (\ref{ca2-forward}) is used to calculate the second-order derivative, and values from two cells (points)
are required. Since a grid is distributed into many \rev{processors}, communications are required to send and receive
information from other \rev{processors}.
On the other hand, these equations show that a communication pattern is also determined by grid distribution and
numerical methods.

\begin{equation}
f'(x) = \frac{f(x + h) - f(x)}{h},
\label{fo-forward}
\end{equation}

\begin{equation}
f'(x) = \frac{f(x) - f(x - h)}{h},
\label{ba-forward}
\end{equation}

\begin{equation}
f'(x) = \frac{f(x + h) - f(x - h)}{2h},
\label{ca-forward}
\end{equation}

\begin{equation}
f''(x) = \frac{f(x + h) + f(x - h) - 2 f(x)}{h^2}.
\label{ca2-forward}
\end{equation}

Equations (\ref{fo-forward}) - (\ref{ca2-forward})
show that a cell requires
information from direct neighbours only. In a parallel computing setting,
each cell is owned by one \rev{processor}, and communications are required to send and receive
information.  However, some numerical methods may need information from
remote neighbours, such as neighbours of a direct neighbour. In this case, \rev{the} communication volume is higher
and its pattern is more complex.

\begin{figure}[!htb]
    \centering
    \includegraphics[width=0.7\linewidth]{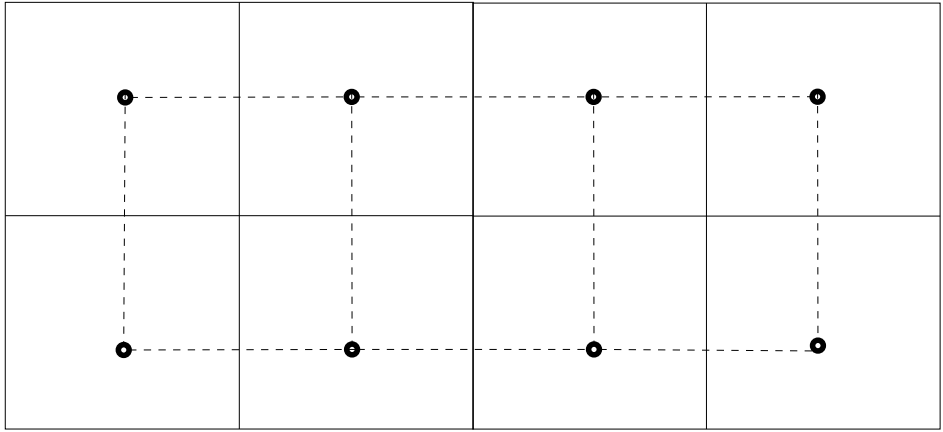}
    \caption{Two-dimensional structured grid and its dual graph}
    \label{fig-dual2d}
\end{figure}

For a certain grid and a numerical method, their communication pattern can be modeled by a
graph theory. If only direct neighbours are involved, then a dual graph can model their
communication pattern. However, if remote neighbours are included, a
pattern is more complex and \rev{a hyper graph, a generalized graph whose edge can
join any number of vertices, must be employed \cite{parmetis}}.
A communication pattern or communication graph can be constructed this way:
\rev{Each cell is read as a vertex, and if two cells
have information exchange, there exists an edge between these two vertices}.
\rev{Figure \ref{fig-dual2d} shows a simple two-dimensional grid and its dual graph,}
which can model a communication pattern of numerical methods that require information from direct
neighbours only.

Each cell may have different workloads, and a weight can be assigned to \rev{represent} a workload. A
communication volume may be different, too, and a weight can be assigned to \rev{represent}
a communication volume between two cells. If all computational nodes have the same computing
capacity, then the objectives of grid partitioning are: 1) all computational nodes (processors
or sub-grids) have an equal or similar workload; 2) communications between them are minimized. The total
communications equal the sum of weights of all edges that are cut.

Graph-based methods are excellent for grid partitioning, which can generate optimal
or quasi-optimal partitionings \cite{nc12,parmetis}. As mentioned above, many graph methods have been
proposed, such as spectral methods \cite{nc2,nc9}, multilevel methods \cite{nc10,nc12}
and diffusive methods \cite{zoltan-ug, zoltan-dev}. \rev{
    Several graph partitioning packages have been implemented
and available publicly, such as METIS \cite{metis}, ParMETIS \cite{parmetis},
Scotch \cite{scotch}, PTScotch \cite{ptscotch}, Chaco \cite{chaco} and Jostle \cite{jostle}.
}

\section{{Space-filling Curves and Orders}}
\label{sec-sfc}

Space-filling curves are those curves that fill an entire $n$-dimensional unit hypercube,
which were proposed by Peano in 1890 and popularized by Hilbert later \cite{Hans, chenyang}.

\begin{figure}[!htb]
    \centering
    \includegraphics[width=.9\linewidth]{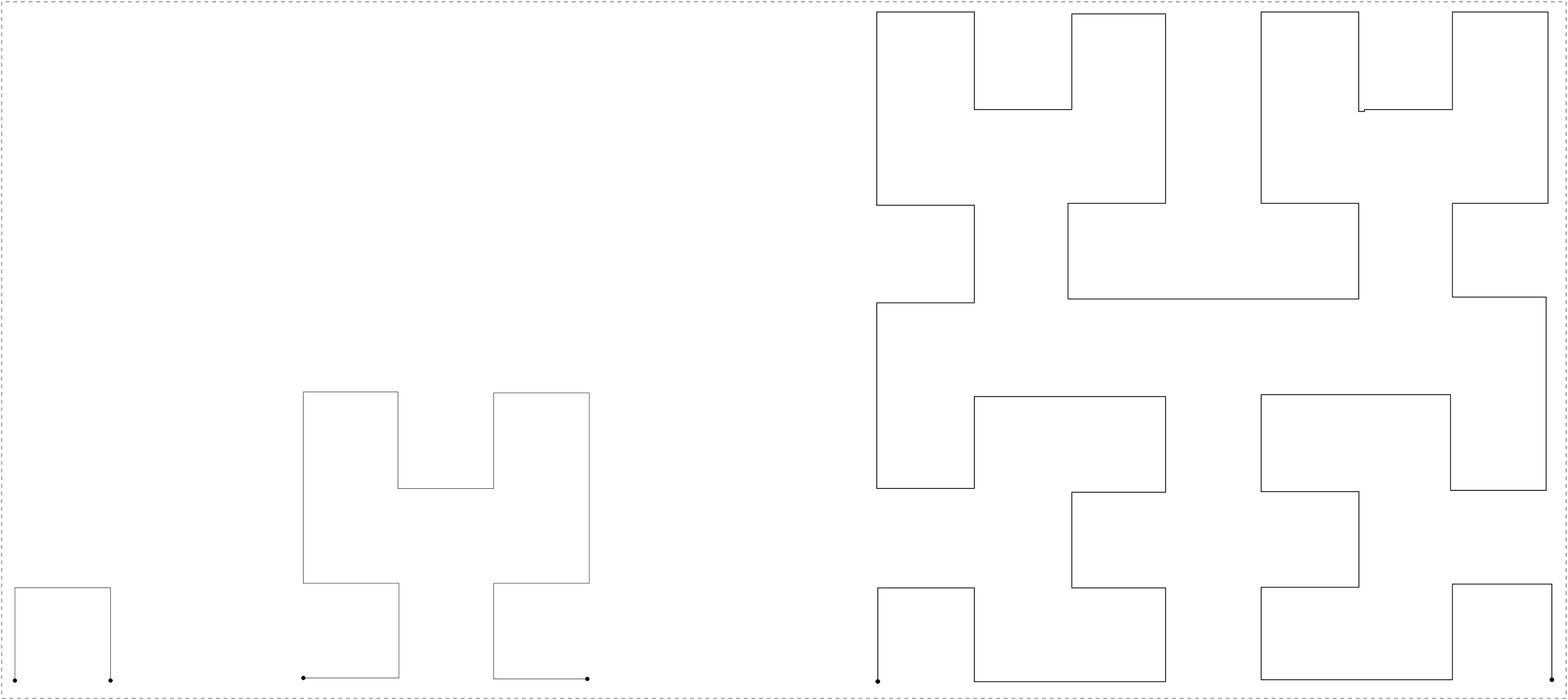}
    \caption{Hilbert space-filling curves, two dimensions, levels 1, 2 and 3}
    \label{fig-hilbert}
\end{figure}

\begin{figure}[!htb]
\begin{center}
\includegraphics[width=0.5 \hsize,angle=270]{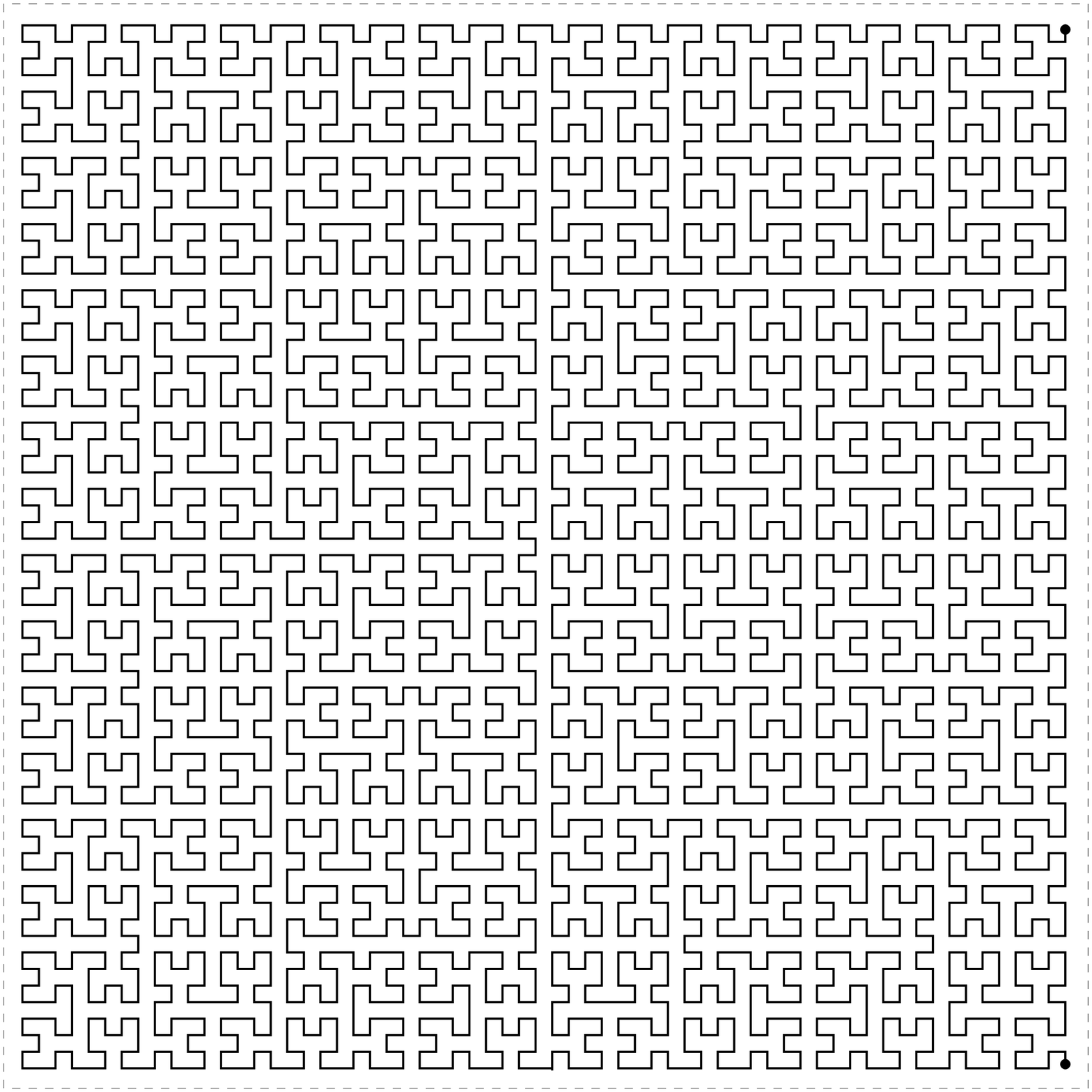}
\caption{Hilbert space-filling curve, level 6}
\label{fig-sfc-h6}
\end{center}
\end{figure}

\begin{figure}[!htb]
\begin{center}
\includegraphics[width=1.\hsize]{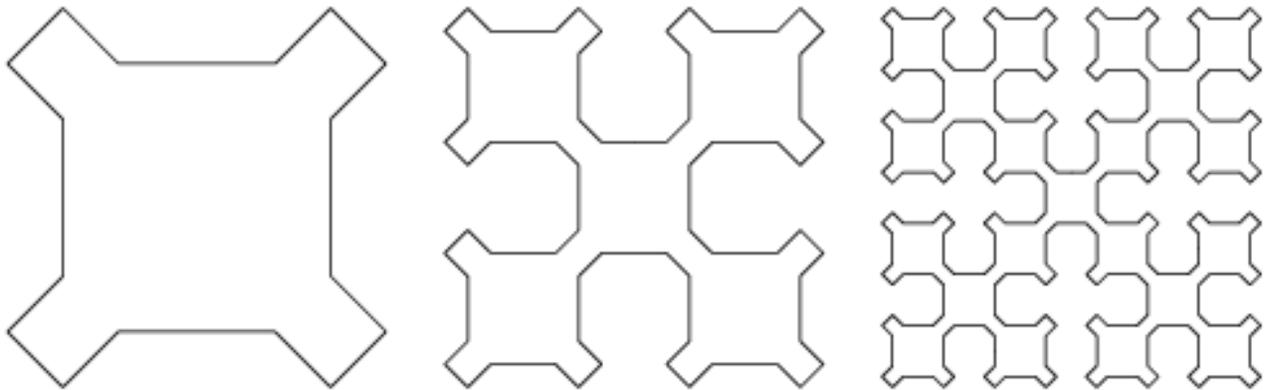}
\caption{Sierpi\'{n}ski space-filling curves, levels 1, 2 and 3}
    \label{fig-sfc-sie}
\end{center}
\end{figure}

\begin{figure}[!htb]
\begin{center}
\includegraphics[width=0.56 \hsize]{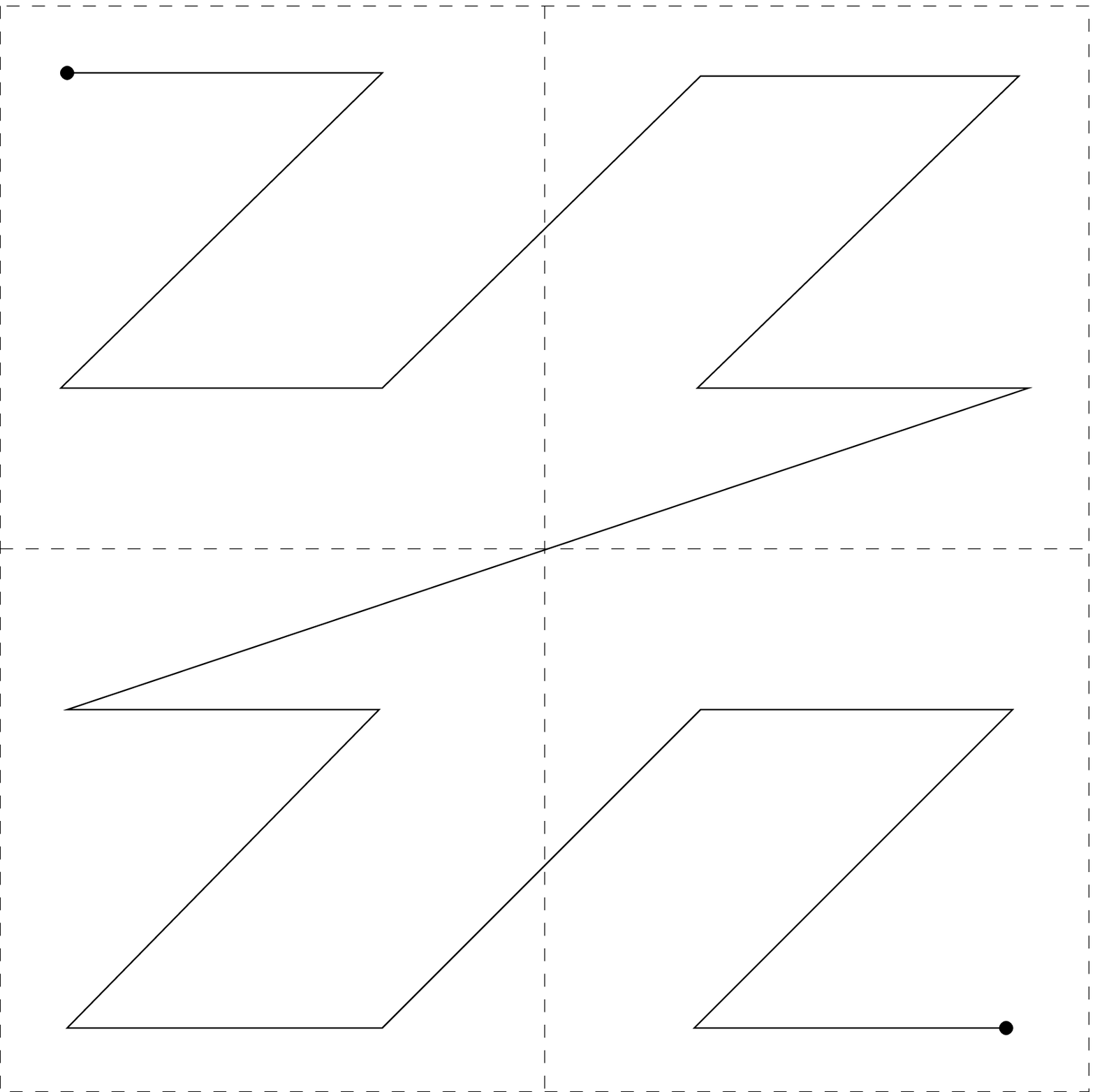}
\caption{Morton space-filling curve, level 2}
    \label{fig-sfc-z}
\end{center}
\end{figure}

Many space-filling curves have been discovered.
Figures \ref{fig-hilbert} and \ref{fig-sfc-h6} show levels 1, 2, 3 and 6 Hilbert space-filling curves
in a two-dimensional unit square. \rev{It can be seen} that a curve is denser if a level is higher.
Figure \ref{fig-sfc-sie} shows levels 1, 2 and 3 Sierpi\'{n}ski space-filling curves.
Figure \ref{fig-sfc-z} shows a level 2 Morton space-filling curve.
\rev{These curves show that} the Hilbert space-filling curves and the Sierpi\'{n}ski space-filling curves
have good locality \rev{(nearby points in the multiple dimensional space are mapped to nearby points
in one-dimensional space \cite{Hans, hsfc-ene, hsfc-cluster})},
and the Morton space-filling curves have jumps, whose locality is poor.

\subsection{{Hilbert Space-filling Curve Orders}}

\begin{figure}[!htb]
\begin{center}
\includegraphics[width=0.5 \hsize]{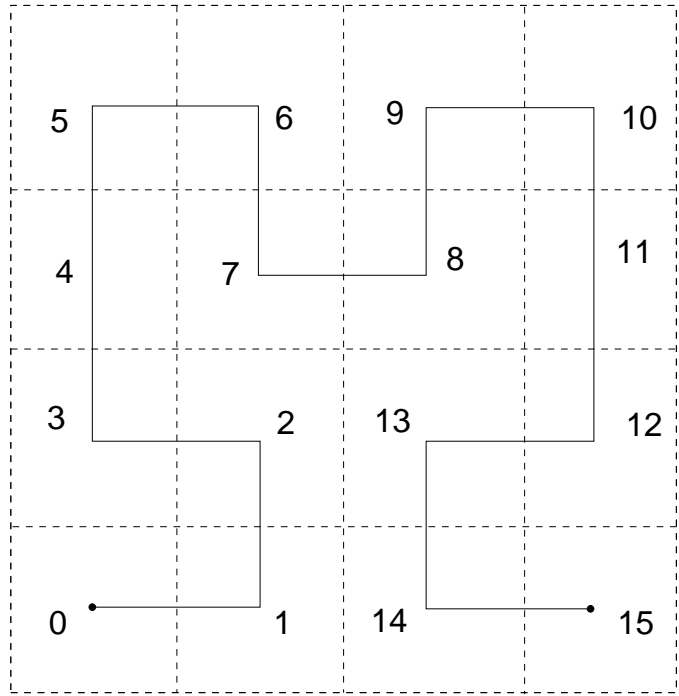}
    \caption{\rev{Hilbert space-filling curve order, level 2}} \label{fig-sfc-h2o}
\end{center}
\end{figure}

\begin{figure}[!htb]
\begin{center}
\includegraphics[width=0.5 \hsize]{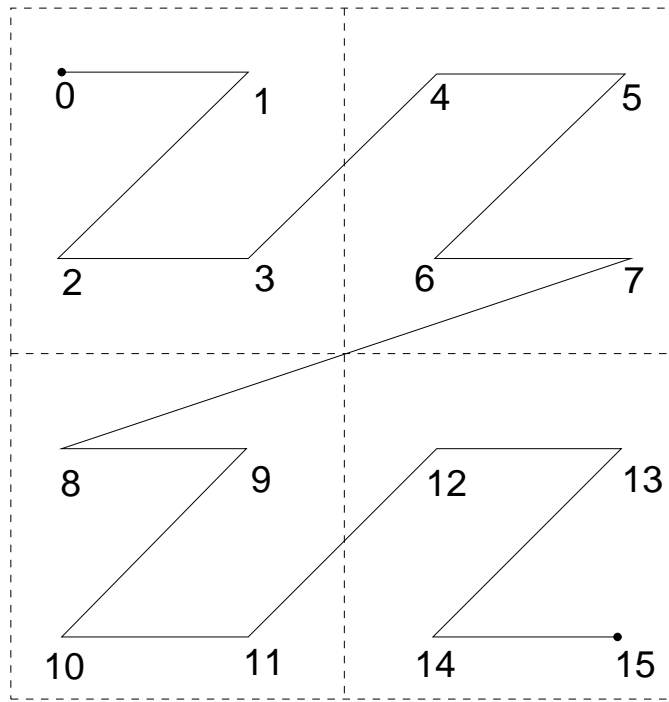}
    \caption{\rev{Morton space-filling curve order (Z-order), level 2}} \label{fig-sfc-zo}
\end{center}
\end{figure}

Each curve has a starting point and an ending point. Along this curve,
a map is introduced between a one-dimensional domain and a multi-dimensional domain.
Figures \ref{fig-sfc-h2o} and \ref{fig-sfc-zo} show a level 2 Hilbert space-filling curve and a level 2
Morton space-fill curve, respectively.
Both of them have 16 vertices whose indexes start from 0 to 15,
\rev{which means that these curves define orders and they map a two-dimensional space
to a one-dimensional space}.
Higher dimensional space-filling curves are defined similarly.

\rev{A Hilbert space-filling curve \cite{Hans} has many important characteristics, such as
locality
and self-similarity (an object is self-similar if small portions of it are reproductions
of initial objects \cite{hsfc-self}).}
A Hilbert curve (order) has been applied in many areas,
including image storing, database indexing, data compression and load balancing.
For parallel computing, the Hilbert order method is one of the most important
geometry-based partitioning methods.

\rev{
    Algorithms for computing Hilbert curves and Hilbert space-filling curve orders
}
in two- and three-dimensional spaces have been proposed in
the literature, which can be classified into recursive algorithms
\cite{butz, gold, witten, cole} and iterative algorithms \cite{griff, sfc-tr-03, xliu2, xliu, fisher, ningtao}.
Iterative algorithms, especially the table-driven algorithms \cite{griff,
sfc-tr-03}, are usually much faster than recursive algorithms.
In general, the complexities of these algorithms are $O(m)$, where $m$ is
the level of a Hilbert curve. For a two-dimensional space,
Chen et al. \cite{ningtao} proposed an algorithm of $O(r)$ complexity,
where $r$ is defined as $r=\log_2(\max(x,y))+1$, $r \le m$ and is independent of the level $m$.
This algorithm is faster when $m$ is much larger than $r$.
The same idea was also applied to a three-dimensional space \cite{HSFC}.
For higher dimensional spaces, Kamata {et al.} presented a representative $n$-dimensional Hilbert
mapping algorithm \cite{kamata} and Li {et al.} introduced algorithms for analyzing the properties of
$n$-dimensional Hilbert curves \cite{chenyang}. Liu et al. introduced high-order encoding
and decoding algorithms \cite{liuh}.

\subsection{Calculation of Hilbert Space-filling Curve Orders}

Table-driven algorithms were introduced in \cite{griff, sfc-tr-03}. The basic idea is
to store additional information other than to calculate. More memory
is required, but computations are faster.

In \cite{sfc-tr-03}, the authors introduced the Gray coding and Morton ordering, which are
easy to compute. Other orderings, such as the
Hilbert ordering, can be generated through pre-defined appropriate mappings. An
ordering table and an orientation table are required to map between the Hilbert and Morton orders.
\rev{
    The appendix presents an algorithm for the calculation of the Hilbert order in Figures \ref{ho-sttable}
    and \ref{ho-algtable}. The code is written in C and can be compiled
    by any C compiler.
    Figure \ref{ho-sttable} shows basic data structures, data types, an
    ordering table, an orientation table, and internal state conversion rules.
    Figure \ref{ho-algtable} is the code for generating a Hilbert order in a three-dimensional
    space, which maps $[0, 1]^3$ to $[0, 1]$.
    \textit{We remark that}
    \begin{itemize}
        \item Here double precision is applied to floating-point numbers and \texttt{int}
            is applied to integers, which can be modified to long double precision and
            longer integer types, such as \texttt{long long}.

        \item \texttt{HSFC\_ENTRY} consists of a three-dimensional coordinate and a value.
            The value is calculated by the algorithm.

        \item \text{hsfc\_maxlevel} defines the maximal level of the Hilbert space-filling
            curve, in which a level of 30 is applied to 32-bit integers. If 64-bit integers are employed,
            a higher level can be applied, such as 61 to increase the number of orders.

        \item \texttt{idata3d}, \texttt{istate3d}, \texttt{d} and \texttt{s} are internal
            tables for Hilbert space-filling curves.

    \end{itemize}
}

\section{Space-filling Curve Partitioning Method}

\label{sec-sfcp}

\rev{
    A space-filling curve defines a map from a multi-dimensional space to a one-dimensional space.
    Since Hilbert space-filling curves have excellent locality, they are good alternatives
    for load balancing methods. Algorithm \ref{alg-hsfc} shows the process of the space-filling
    curve methods, which has three steps:
    \begin{enumerate}
        \item The first step is to map the given computational domain $\Omega$ to a subset of $(0, 1)^3$
            (boundaries are excluded), which can be obtained by a linear mapping.
            In this case, for any cell, the coordinate used to represent the cell belongs to $(0, 1)^3$.
            In Algorithm \ref{alg-hsfc}, when the computational domain is mapped,
            its aspect ratio is preserved.

        \item A space-filling curve is employed to convert coordinates from $(0, 1)^3$ to a value in
            set $(0, 1)$.

        \item The third step is to partition $(0, 1)$ into $N_p$ sub-intervals such that each
            sub-interval has the same number of cells (or workload).
    \end{enumerate}
}

\rev{Each MPI process can perform the first and the second steps
independently. In our implementation, the third step is \rev{calculated} by one MPI process and it
broadcasts results to other MPI  processes. Also, different space-filling curves define different
partitioning methods.
The space-filling curve methods}
assume that two cells (elements) that are close to each other
have a higher possibility to communicate with each other than two cells that are far from each other.
This assumption is true for grid-based numerical methods,
such as the finite element, finite volume and finite difference methods.
The only difference for various space-filling curve methods is how to map a cell to $(0, 1)$.

\begin{algorithm}[!htb]
\caption{Space-filling curve method}
\label{alg-hsfc}
\begin{algorithmic}[1]
\STATE Map a computational domain $\Omega$ to a subset of $(0, 1)^3$.
\STATE For any cell, calculate its mapping that belongs to $(0, 1)$.
    \STATE Partition the interval $(0, 1)$ into $N_p$ sub-intervals \rev{such that} each sub-interval has the same number of cells.
\end{algorithmic}
\end{algorithm}

\subsection{Partition Quality}

This section studies partition quality on homogeneous architectures.
The quality of a partition resulting from the graph methods and space-filling curve methods
has several measurements, including a local surface index
and a global surface index. These concepts have been studied in \cite{liuh}.

Let $f_i$ be the number of faces in the \rev{$i$th} sub-grid and $b_i$ be the number of faces
that are shared by a cell in another \rev{process}.
$b_i$ is proportional to the communication volume of the $i$th \rev{process} involved.

The maximum local surface index is defined by
\begin{equation}
    r_M = \max_{0\le i < \rev{N_p}}\frac{b_i}{f_i}.
\end{equation}
It is used to model the maximal communications that one \rev{process} involves.

The average surface index is defined by
\begin{equation}
    r_A = \sum_{i = 0}^{N_p - 1}\frac{b_i}{f_i}.
\end{equation}
It is used to model the average communications that one \rev{process} involves.

If a cell in a sub-grid has a neighbour in another \rev{processor}, then these two
sub-grids are connected. Inter-\rev{processors} connectivity is the number of connected sub-grids of
\rev{a given sub-grid assigned to a processor}, denoted by $c_i$. The maximal inter-\rev{processors} connectivity is defined by
\begin{equation}
c = {\max_{0 \le i < N_p} c_i}.
\end{equation}

\section{Numerical Experiments}
\label{sec-num}

Several cases are studied in this section. \rev{GPC (General Purpose Cluster)} from SciNet and
Blue Gene/Q from IBM US
are employed for numerical experiments. The GPC is a regular cluster that uses
Intel processors.
The Blue Gene/Q has two racks, and each rack has two midplanes.
Each midplane contains 16 computer nodes.
Each computer node has 32 computer cards (64-bit PowerPC A2 processor) and one computer card has 17 cores,
one of which is responsible for the operating system and the other
16 cores for computation.
The IBM Blue Gene/Q system has a total of 32,768 CPU cores for computation.

\rev{
    This section has three subsections. The first subsection focuses on partitioning quality,
    and the ParMETIS and HSFC (Hilbert space-filling curve) methods are compared. The second
    subsection compares performance of reservoir simulations using the ParMETIS and HSFC methods.
    The last subsection presents the scalability of parallel reservoir simulations using the HSFC method.

    The reservoir simulator that is used to study the load balancing is based on our parallel
    in-house platform \cite{liuh}. Structured grids are employed and cell-centered data
    is applied to represent a reservoir, such as porosity, permeability, pressure and saturation.
    The discretization method is the finite difference (or volume) method with upstream weighting techniques.
    The linear solvers are Krylov subspace solvers, such as GMRES(m), LGMRES(m) and BICGSTAB.
    The preconditioners are multi-stage methods \cite{bos-pc,CEL},
    which are combinations of the AMG methods for pressure and the RAS (Restricted Additive
    Schwarz) method \cite{RAS}, one of the domain decomposition methods. The nonlinear methods are
    the standard Newton method and the inexact Newton method, where the only difference is how
    to choose termination tolerance. The Peaceman model \cite{PWM} is applied for well management.
    When a grid is distributed, the destination MPI process of each grid cell
    is determined by a load balancing
    algorithm, so the well segments of a well may exist in many MPI processes.
    A map is defined, which manages the global numbering of unknowns, their local numbering
    on each MPI process, and the numbering of their neighbours.

    In reservoir simulation, when a grid is given and numerical methods are chosen, the scalability
    of the reservoir simulator is determined by communications, most of which happen in the solution
    procedure of linear systems. The distribution of a grid and a linear system also affects the performance
    of linear solvers.

}

\subsection{Partitioning Quality}


\begin{example}
    \label{dlb-ex2}
    The computational domain is $1200 ft \times 2200 ft \times 170 ft$ and the grid
    size is $180 \times 660 \times 255$, which has 30,294,000 cells. \rev{The grid is a refinement
    of the grid from SPE10 project \cite{SPE10}.} The HSFC method
    and ParMETIS are applied to test partitioning quality. Numerical results are shown
    in Tables \ref{tab-dlb2-si} and \ref{tab-dlb2-ic}.
\end{example}

\begin{table}[!htb]
    \centering \tabcolsep=5pt \caption{Surface indices for \rev{Example} \ref{dlb-ex2}}
\label{tab-dlb2-si}
    \begin{tabular}{|l|c|c|c|c|} \hline
        \multicolumn{5}{|c|}{maximum surface index ($r_M$,\%)} \\ \hline
        \# \rev{MPI ranks} &256    & 512   & 1024   & 2048  \\ \hline
        ParMETIS     & 6.86 & 9.23 & 11.0 & 13.6 \\
        HSFC         & 8.34 & 10.7 & 13.5 & 18.2  \\

        \hline
        \multicolumn{5}{|c|}{average surface index ($r_A$,\%)} \\ \hline
        \# \rev{MPI ranks} &256    & 512   & 1024   & 2048  \\ \hline
        ParMETIS     & 5.15 & 6.55 & 8.37 & 10.5  \\
        HSFC         & 6.39 & 8.16 & 10.4 & 13.0  \\

        \hline
    \end{tabular}
\end{table}

\begin{table}[!htb]
    \centering \caption{Maximal inter-\rev{processors} connectivities ($c$), Example \ref{dlb-ex2}}
    \label{tab-dlb2-ic}
    \begin{tabular}{|l|c|c|c|c|} \hline
        \# \rev{MPI ranks} &256    & 512   & 1024   & 2048  \\ \hline
        ParMETIS     & 19  & 20  & 21 & 26  \\
        HSFC         & 16  & 15  & 15 & 16  \\
        \hline
    \end{tabular}
\end{table}

\rev{
    Table \ref{tab-dlb2-si} compares the maximum surface index and average surface index, which model
    the communication volume. A smaller value indicates less communication and better partitioning quality.
    The table shows that communications resulted from ParMETIS are less than those from the HSFC
    method, which shows that the graph-based methods can minimize communications.
    Table \ref{tab-dlb2-ic} compares the inter-\rev{processors} connectivity.
    However, in this case, the HSFC method has a better inter-\rev{process} connectivity than ParMETIS,
    and the resulting partitioning has less latency.
}

\subsection{Numerical Performance}
\rev{
    This section studies the numerical performance of reservoir simulations using the ParMETIS
    and HSFC methods.
}

\begin{example}
    \label{dlb-ogw}
    This case tests a three-phase black oil problem with a refined SPE10 geological model,
    where each cell is refined into \rev{8} smaller cells. The model
    has around \rev{9} millions of cells. The nonlinear method is an inexact Newton method.
    The solver is GMRES(30) and the preconditioner is the CPR-PF method \cite{bos-pc}.
    \rev{200} time steps are applied. The case is run on GPC.
    Numerical summaries are shown in Table \ref{tab-ogw-num}.
\end{example}

\begin{table}[!htb]
    \centering \tabcolsep=5pt \caption{Numerical summaries for \rev{Example} \ref{dlb-ogw}}
    \label{tab-ogw-num}
    \begin{tabular}{|l|c|c|c|c|c|} \hline
        \multicolumn{6}{|c|}{\# Newton iterations}  \\ \hline
        \# \rev{MPI ranks}      & 64    & 128   & 256   & 512  & 1024 \\ \hline
        ParMETIS     & 769   & 785   & 783   & 762  & 763 \\
        HSFC         & 742   & 746   & 730   & 747  & 742 \\

        \hline
        \multicolumn{6}{|c|}{\# Linear iterations}   \\ \hline
        \# \rev{MPI ranks}      & 64    & 128   & 256   & 512  & 1024 \\ \hline
        ParMETIS     & 18177  & 21162  & 19736  & 20297 & 19103 \\
        HSFC         & 16971  & 16072  & 18210  & 19021 & 22712 \\

        \hline

        \multicolumn{6}{|c|}{Overall running time (s)} \\ \hline
        \# \rev{MPI ranks}      & 64    & 128   & 256   & 512  & 1024 \\ \hline
        ParMETIS     & 24529.75 & 13636.87 & 6449.05 & 3518.78 & 2604.95 \\
        HSFC         & 23486.31 & 12108.27 & 6440.79 & 3858.01 & 3100.40  \\

        \hline
    \end{tabular}
\end{table}

\rev{
    The CPR-PF method is a two-stage preconditioning method designed for the black oil and
    compositional models, which is a scalable method proposed for parallel reservoir simulation.
    The first stage is to solve the pressure system using a
    parallel algebraic multi-grid (AMG) method, which is represented by \texttt{P}.
    The second stage is to solve the entire system using the restricted additive Schwarz (RAS)
    method \cite{bos-pc}, one of the domain decomposition methods,
    which is represented by \texttt{F}.
    For parallel algebraic multi-grid methods, interior nodes and boundary nodes are handled
    differently to increase parallelism and to reduce complexity, and a compromise must be made
    when generating interpolation operators to reduce complexity and communications. In addition,
    for a domain decomposition method, a smaller sub-problem needs to be solved in each MPI
    process. In another word, the grid distribution and the resulted matrix distribution affect
    the performance of linear solvers and preconditioners and the performance of a nonlinear method.

    Table \ref{tab-ogw-num} shows the number of Newton iterations, the number of linear iterations
    and overall running time. For the Newton method, it is clear that the simulations with the HSFC partitioning
    method always use fewer iterations than the simulations with ParMETIS, which indicates that the
    nonlinear method with HSFC has better convergence. The simulations with the
    HSFC partitioning method use fewer linear iterations than the simulations with ParMETIS except
    that 1024 MPI processes are employed. It is evident that the partitioning methods affect the behavior
    of the numerical simulations, and the simulations with the HSFC partitioning method are more
    robust. Considering running time, the simulations with the HSFC partitioning
    method outperform the simulations with ParMETIS using up to 256 MPI processes. When 512 
    and 1024 MPI processes are employed, the simulations with ParMETIS run faster than the simulations with
    HSFC, even though the latter has fewer Newton iterations.
    The difference is caused by communications in the linear solver, which shows again that the graph-based
    methods have better partitioning quality than the geometric methods.
}

\begin{example}
    \label{dlb-ow}
    This case tests a two-phase oil-water model with a refined SPE10 project,
    where each original cell is refined into 27 smaller cells. The model
    has around 30 millions of cells. The nonlinear method is an inexact Newton method.
    \rev{The solver is GMRES(50), which is the restarted generalized minimal residual method (GMRES).
    The preconditioner is the CPR-PF method \cite{bos-pc}.
    }
    20 time steps are applied. The case is run on GPC.
    Numerical summaries are shown in Table \ref{tab-ow-num}.
\end{example}

\begin{table}[!htb]
    \centering \tabcolsep=5pt
    \caption{Numerical summaries for \rev{Example} \ref{dlb-ow}}
    \label{tab-ow-num}
    \begin{tabular}{|l|c|c|c|c|} \hline
        \multicolumn{5}{|c|}{\# Newton iterations} \\ \hline
        \# \rev{MPI ranks} &256    & 512   & 1024   & 2048  \\ \hline
        ParMETIS     & 106 & 105 & 105 & 79 \\
        HSFC         & 101 & 80 & 105 & 101  \\

        \hline
        \multicolumn{5}{|c|}{\# Linear iterations} \\ \hline
        \# \rev{MPI ranks} &256    & 512   & 1024   & 2048  \\ \hline
        ParMETIS     & 364 & 231 & 332 & 375  \\
        HSFC         & 251 & 289 & 379 & 296  \\

        \hline

        \multicolumn{5}{|c|}{Overall running time (s)} \\ \hline
        \# \rev{MPI ranks} &256    & 512   & 1024   & 2048  \\ \hline
        ParMETIS     & 668.5 & 346.6 & 236.1 & 116.4  \\
        HSFC         & 621.0 & 288.4 & 295.2 & 187.2  \\

        \hline
    \end{tabular}
\end{table}

\rev{
Table \ref{tab-ow-num} shows numerical summaries, including the Newton method, linear solver and overall
running time. For the Newton method, when ParMETIS is applied, the number of nonlinear
iterations is similar using
up to 1024 MPI processes. However, when 2048 MPI processes are employed, a sudden reduction is observed.
When the HSFC method is applied, the number of nonlinear iterations is similar except the case of
512 MPI processes. The HSFC method uses fewer Newton iterations for the cases of 256 MPI processes and
512 MPI processes. The ParMETIS outperforms the HSFC method when 2048 MPI processes are employed.

Regarding the linear solver, the simulations using HSFC use fewer iterations than those with ParMETIS when
256 MPI processes and 2048 MPI processes are employed. The ParMETIS has better performance when
512 and 1024 MPI processes are employed.

Considering the overall running time, Table \ref{tab-ow-num} shows that when 256 and 512 MPI
processes are used, the simulations with HSFC are faster than the simulations using
ParMETIS. When using 1024 and 2048 MPI processes, the simulations with ParMETIS are faster.

It is well-known that ParMETIS is an ideal tool for load balancing. However, in this example, the
results show that the HSFC method can outperform ParMETIS in some cases, and the HSFC method
is a good alternative for parallel reservoir simulations.
}

\subsection{Scalability}

\rev{
    This section studies the scalability of reservoir simulations using the HSFC method.
}

\begin{figure}[!htb]
    \centering
    \includegraphics[width=0.5\linewidth, angle=270]{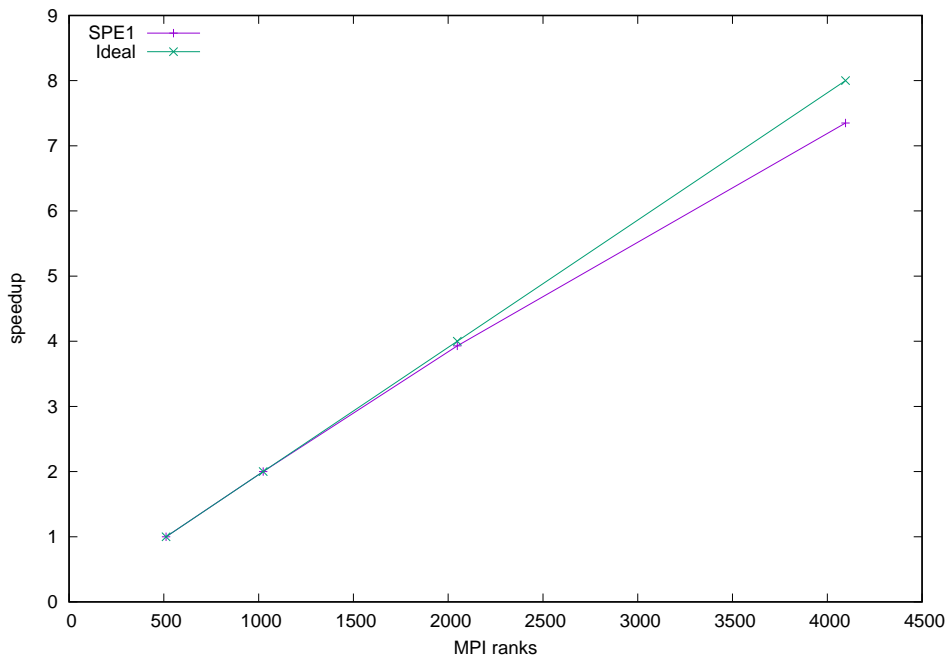}
    \caption{Scalability of Example \ref{bos-spe1-bgq}, overall}
    \label{fig-bos-spe1-bgq}
\end{figure}

\begin{example}
\label{bos-spe1-bgq}
    This example tests the black oil model using a refined SPE1 problem,
    which has 100 millions of grid cells. The nonlinear method is an
    inexact Newton method. The linear solver is BiCGSTAB, and the preconditioner
    is the CPR-PF method. The simulation time is 10 days. The experiments are
    run on IBM Blue Gene/Q.  The grid partitioning method is the HSFC method.
    Numerical summaries are shown in Table \ref{tab-bos-spe1-bgq}
    and scalability is presented in Figure \ref{fig-bos-spe1-bgq}.
\end{example}

\begin{table}[!htb]
    \centering
    \caption{Numerical summaries of Example \ref{bos-spe1-bgq}}
    \begin{tabular}{|c|c|c|c|c|} \hline
        \# \rev{MPI ranks}   & \# Steps & \# Newton & \# Solver  & Time (s)\\ \hline
        512  & 27 & 126 & 383 & 10606.83 \\
        1024 & 27 & 129 & 377 & 5328.46 \\
        2048 & 26 & 122 & 362 & 2703.51 \\
        4096 & 27 & 129 & 394 & 1474.21 \\
        \hline
    \end{tabular}
    \label{tab-bos-spe1-bgq}
\end{table}

\rev{
    This case uses up to 4096 MPI processes.
    Table \ref{tab-bos-spe1-bgq} shows time steps, the number of Newton iterations and the number
    of linear iterations. In this example, the Newton method and linear solver show
    similar convergence using different numbers of MPI processes.
    Figure \ref{fig-bos-spe1-bgq} presents a strong scalability curve,
    and it indicates that our simulator, which uses the HSFC method to partition a grid,
    has excellent scalability, and the HSFC method is a good partitioning method for
    parallel reservoir simulations.
}

\begin{figure}[!htb]
    \centering
    \includegraphics[width=0.5\linewidth, angle=270]{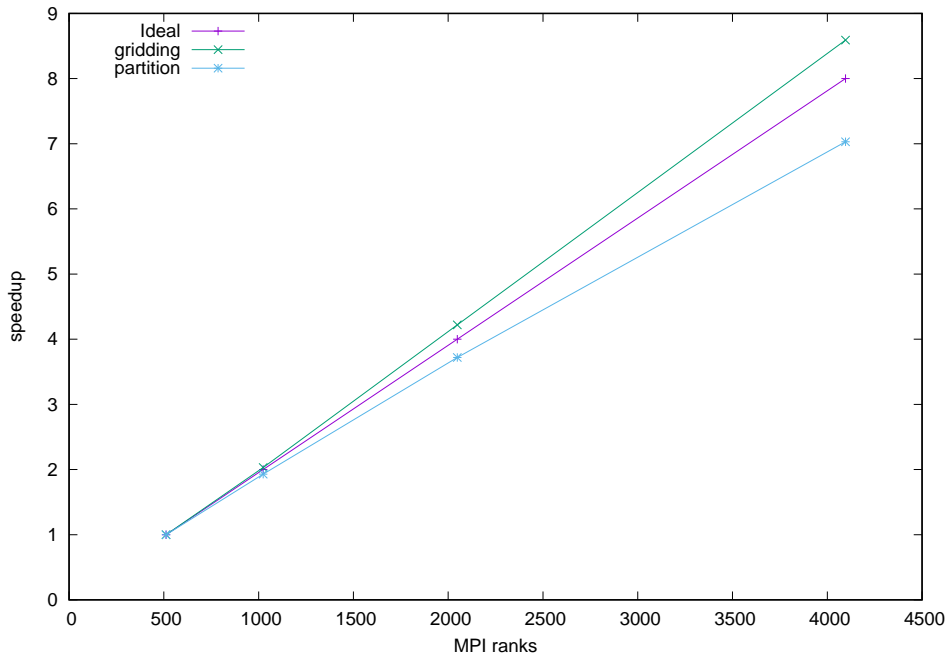}
    \caption{Scalability of Example \ref{po-bgq}, including partitioning and gridding}
    \label{fig-po-bgq-grid}
\end{figure}

\begin{figure}[!htb]
    \centering
    \includegraphics[width=0.5\linewidth, angle=270]{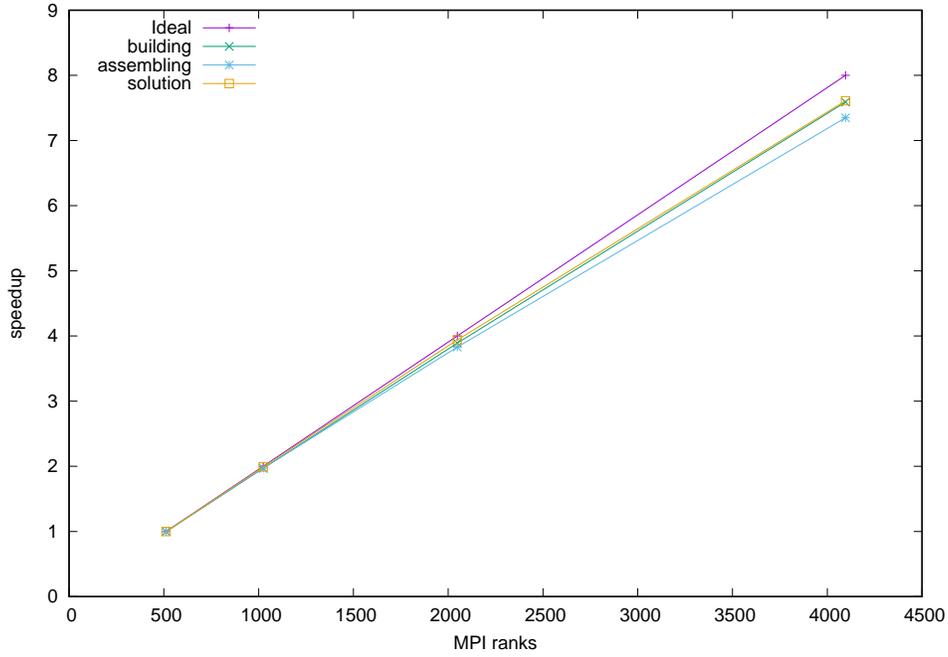}
    \caption{Scalability of Example \ref{po-bgq}, including building, assembling and solution}
    \label{fig-po-bgq-solver}
\end{figure}

\begin{figure}[!htb]
    \centering
    \includegraphics[width=0.5\linewidth, angle=270]{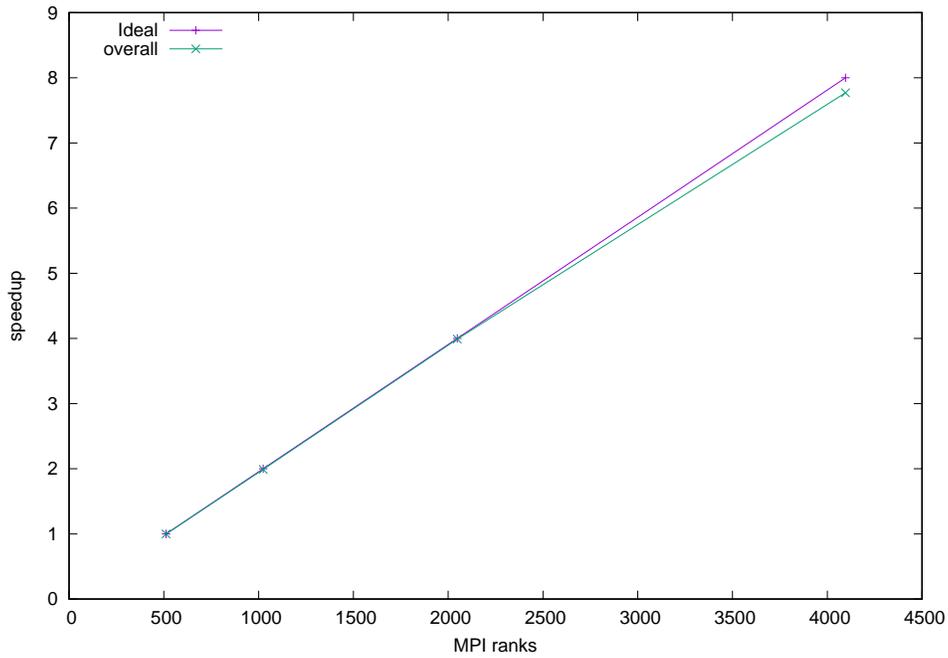}
    \caption{Scalability of Example \ref{po-bgq}, overall}
    \label{fig-po-bgq}
\end{figure}

\begin{example}
\label{po-bgq}
    This example tests a linear system from a \rev{Poisson} equation. The grid size
    is five billion. The GMRES(30) solver is employed and it is set to run 90 iterations.
    The preconditioner is the RAS (Restricted Additive Schwarz) method.
    The experiments are run on IBM Blue Gene/Q. Again, the grid partitioning method
    is the HSFC method. Numerical summaries are shown in Table \ref{tab-po-bgq}
    and the scalability curves for partitioning, gridding, the linear solver
    and overall run time are shown in Figures \ref{fig-po-bgq-grid} -
    \ref{fig-po-bgq}.
\end{example}

\begin{table}[!htb]
\centering
  \caption{Numerical summaries of Example \ref{po-bgq}}
\begin{tabular}{|c|c|c|c|c|c|c|} \hline
    \# \rev{MPI ranks}  & Partitioning (s)& Gridding (s)& Build (s)& Assemble (s)& Solve (s) & Overall time (s) \\ \hline
    512  & 23.12 & 337.09 & 48.12 & 109.07 & 998.77 & 1504.32 \\
    1024 & 11.98 & 166.21 & 24.35 & 55.29 & 502.92 & 754.68 \\
    2048 & 6.21  & 79.92  & 12.38 & 28.46 & 253.46 & 377.35 \\
    4096 & 3.29  & 39.26  & 6.34  & 14.83 & 131.28 & 193.52 \\
 \hline
\end{tabular}
  \label{tab-po-bgq}
\end{table}

\rev{
    This example tests the scalability of a simple \rev{Poisson} equation, and
    this problem has fixed calculations, which is another proper case to
    study scalability of parallel applications. Even though it is a Poisson equation,
    its calculations are similar to reservoir simulations, including building linear systems,
    assembling linear solvers and preconditioners, and solution of the linear systems.
}
Numerical summaries are shown in Table \ref{tab-po-bgq}, including the number of \rev{MPI ranks},
partitioning time, grid generating time, linear system building time,
solver assembling time, solution time and overall running time.
The running time for partitioning and Figure \ref{fig-po-bgq-grid} show that
the partitioning method is slightly worse than the ideal case. The reason is that
the third step (one-dimensional partitioning) of the partitioning method
is performed by one MPI and it broadcasts results to other MPIs,
which limits the scalability of the method. However, other modules, such as
gridding, the linear solver (building, assembling and solution) and the preconditioner
have excellent scalability, which are demonstrated in Figures \ref{fig-po-bgq-solver}
and \ref{fig-po-bgq}.
\rev{
    This example shows that our parallel application
}
with the HSFC partitioning
method has excellent scalability and the HSFC method is a good choice for
large-scale simulations.

\section{Conclusion}
This paper presents our work on developing the Hilbert space-filling curve (HSFC) based
grid partitioning method,
\rev{which consists of three steps. The first step is to map the computational domain
to a cube.  The second step is to convert the three-dimensional space to one-dimensional space, $(0,1)$,
using a mapping defined by a Hilbert space-filling curve. The third step is to partition the
one-dimensional space to $N_p$ sub-intervals, where $N_p$ is the number of MPI processes.}
Numerical experiments are carried out to compare the HSFC method with
\rev{ParMETIS, a graph based partitioning method.
These experiments show that the HSFC method has good partition quality.
Large-scale simulations indicate that our simulator using the HSFC partitioning
method has excellent scalability and the HSFC partitioning method is a good alternative
to graph methods in scientific computing.}

\section*{Acknowledgements}
The support of Department of Chemical and Petroleum Engineering,
University of Calgary and Reservoir Simulation Group is gratefully
acknowledged. The research is partly supported by NSERC/AIEES/Foundation
CMG, AITF (iCore), IBM Thomas J. Watson Research Center, and the Frank
and Sarah Meyer FCMG Collaboration Centre for Visualization and Simulation.
The research is also enabled in part by support
provided by WestGrid (www.westgrid.ca), SciNet (www.scinethpc.ca)
and Compute Canada Calcul Canada (www.computecanada.ca).

\appendix

\section{Table-driven Algorithms for Three-Dimensional Hilbert Orders}

\begin{figure}[!htb]
\begin{mvb}
typedef double FLOAT;
typedef int    INT;
typedef int    BOOLEAN;

typedef struct HSFC_ENTRY_ {
    FLOAT coord[3];
    FLOAT hsfc;

} HSFC_ENTRY;

static INT hsfc_maxlevel  = 30;
\end{mvb}

\begin{mvb}
static unsigned const int idata3d[] = {
    0, 7, 3, 4, 1, 6, 2, 5, 0, 1, 3, 2, 7, 6, 4, 5, 0, 3, 7, 4, 1, 2, 6, 5, 2, 3, 5, 4, 1, 0, 6, 7,
    4, 5, 3, 2, 7, 6, 0, 1, 4, 7, 3, 0, 5, 6, 2, 1, 6, 7, 5, 4, 1, 0, 2, 3, 0, 1, 7, 6, 3, 2, 4, 5,
    2, 1, 5, 6, 3, 0, 4, 7, 6, 1, 5, 2, 7, 0, 4, 3, 0, 7, 1, 6, 3, 4, 2, 5, 2, 1, 3, 0, 5, 6, 4, 7,
    4, 7, 5, 6, 3, 0, 2, 1, 4, 5, 7, 6, 3, 2, 0, 1, 6, 1, 7, 0, 5, 2, 4, 3, 0, 3, 1, 2, 7, 4, 6, 5,
    2, 3, 1, 0, 5, 4, 6, 7, 6, 7, 1, 0, 5, 4, 2, 3, 2, 5, 1, 6, 3, 4, 0, 7, 4, 3, 7, 0, 5, 2, 6, 1,
    4, 3, 5, 2, 7, 0, 6, 1, 6, 5, 1, 2, 7, 4, 0, 3, 2, 5, 3, 4, 1, 6, 0, 7, 6, 5, 7, 4, 1, 2, 0, 3
};
\end{mvb}

\begin{mvb}
static unsigned const int istate3d[] = {
    1,  6,  3,  4,  2,  5,  0,  0,  0,  7,  8,  1,  9,  4,  5,  1,
    15, 22, 23, 20, 0,  2,  19, 2,  3,  23, 3,  15, 6,  20, 16, 22,
    11, 4,  12, 4,  20, 1,  22, 13, 22, 12, 20, 11, 5,  0,  5,  19,
    17, 0,  6,  21, 3,  9,  6,  2,  10, 1,  14, 13, 11, 7,  12, 7,
    8,  9,  8,  18, 14, 12, 10, 11, 21, 8,  9,  9,  1,  6,  17, 7,
    7,  17, 15, 12, 16, 13, 10, 10, 11, 14, 9,  5,  11, 22, 0,  8,
    18, 5,  12, 10, 19, 8,  12, 20, 8,  13, 19, 7,  5,  13, 18, 4,
    23, 11, 7,  17, 14, 14, 6,  1,  2,  18, 10, 15, 21, 19, 20, 15,
    16, 21, 17, 19, 16, 2,  3,  18, 6,  10, 16, 14, 17, 23, 17, 15,
    18, 18, 21, 8,  17, 7,  13, 16, 3,  4,  13, 16, 19, 19, 2,  5,
    16, 13, 20, 20, 4,  3,  15, 12, 9,  21, 18, 21, 15, 14, 23, 10,
    22, 22, 6,  1,  23, 11, 4,  3,  14, 23, 2,  9,  22, 23, 21, 0
};
\end{mvb}
\begin{mvb}
    static unsigned const int *d[] = {
        idata3d,       idata3d + 8,   idata3d + 16,  idata3d + 24,
        idata3d + 32,  idata3d + 40,  idata3d + 48,  idata3d + 56,
        idata3d + 64,  idata3d + 72,  idata3d + 80,  idata3d + 88,
        idata3d + 96,  idata3d + 104, idata3d + 112, idata3d + 120,
        idata3d + 128, idata3d + 136, idata3d + 144, idata3d + 152,
        idata3d + 160, idata3d + 168, idata3d + 176, idata3d + 184
    };

    static unsigned const int *s[] = {
        istate3d,       istate3d + 8,   istate3d + 16,  istate3d + 24,
        istate3d + 32,  istate3d + 40,  istate3d + 48,  istate3d + 56,
        istate3d + 64,  istate3d + 72,  istate3d + 80,  istate3d + 88,
        istate3d + 96,  istate3d + 104, istate3d + 112, istate3d + 120,
        istate3d + 128, istate3d + 136, istate3d + 144, istate3d + 152,
        istate3d + 160, istate3d + 168, istate3d + 176, istate3d + 184
    };
\end{mvb}
    \caption{Ordering table, orientation table, state tables for Hilbert order}
    \label{ho-sttable}
\end{figure}

\begin{figure}[!htb]
\begin{mvb}
void HilbertInvOrder3d(HSFC_ENTRY *x)
{
    int level, EffLen;
    unsigned int key[3], c[3], temp, stat;
    INT i;

    static unsigned INTMX;
    static unsigned EfBit;
    static BOOLEAN initialized = FALSE;
    static int k0 = 0, k1 = 0, k2 = 0;

    if (!initialized) {
        initialized = TRUE;

        INTMX = 4294967295U;
        EfBit = INTMX >> 2;

        k0 = 60 - hsfc_maxlevel * 3;
        k1 = 30 - hsfc_maxlevel * 3;
        k2 = -hsfc_maxlevel * 3;
    }

    c[0] = (unsigned int)(x[i].coord[0] * (double)INTMX);
    c[1] = (unsigned int)(x[i].coord[1] * (double)INTMX);
    c[2] = (unsigned int)(x[i].coord[2] * (double)INTMX);
    c[1] >>= 1;
    c[2] >>= 2;

    key[0] = key[1] = key[2] = 0;
    stat = 0;
    EffLen = 30;
    for (level = 0; level < hsfc_maxlevel; level++) {
        EffLen--;
        temp = ((c[0] >> EffLen) & 4) | ((c[1] >> EffLen) & 2) | ((c[2] >> EffLen) & 1);

        key[0] = (key[0] << 3) | ((key[1] >> 27) & 7);
        key[1] = (key[1] << 3) | ((key[2] >> 27) & 7);
        key[2] = (key[2] << 3) | *(d[stat] + temp);

        stat = *(s[stat] + temp);
    }

    key[0] = key[0] & EfBit;
    key[1] = key[1] & EfBit;
    key[2] = key[2] & EfBit;

    x[i].hsfc  = ldexp((double)key[2], k2);
    x[i].hsfc += ldexp((double)key[1], k1);
    x[i].hsfc += ldexp((double)key[0], k0);
}
\end{mvb}
    \caption{Computation of Hilbert order}
    \label{ho-algtable}
\end{figure}

\vskip -0.2in

\begin{thebibliography}{10}

\bibitem{HSFC}
{H. Liu}, {Dynamic Load Balancing on Adaptive Unstructured Meshes}, {10th IEEE International Conference on
 High Performance Computing and Communications}, 2008.

\bibitem{bos-pc}
H. Liu, K. Wang, Z. Chen, and K. E. Jordan, Efficient Multi-stage Preconditioners
for Highly Heterogeneous Reservoir Simulations on Parallel Distributed Systems,
SPE-173208-MS, SPE Reservoir Simulation Symposium held in Houston, Texas, USA, 23-25 February 2015.

\bibitem{RAS}
X. Cai, and M. Sarkis,
A restricted additive Schwarz preconditioner for general sparse linear systems,
SIAM Journal on Scientific Computing 21.2 (1999): 792-797.

\bibitem{sfc-tr-03}
P. M. Campbell and K. D. Devine and J. E. Flaherty and L. G. Gervasio and J. D. Teresco,
Dynamic load balancing using space-filling curves, Technical Report CS-03-01, 2003.

\bibitem{mitchell-k}
William F Mitchell, The Refinement-Tree Partition for Parallel Solution of Partial Differential Equations, NIST Journal of Research, 103, 405-414, 1998.

\bibitem{mitchell-hp}
William F Mitchell, Hamiltonian Paths Through Two- and Three-Dimensional Grids, NIST Journal of Research, 110,127-136, 2005.

\bibitem{metis}
George Karypis and Kirk Schloegel and Vipin Kumar, PARMETIS: Parallel Graph Partitioning and Sparse Matrix Ordering Library version 3.1, 2003.

\bibitem{zoltan-ug}
E. Boman and K. Devine and R. Heaphy and B. Hendrickson and V. Leung and L.A. Riesen and C. Vaughan and U. Catalyurek and D. Bozdag and W. Mitchell and J. Teresco, Zoltan v3: Parallel Partitioning, Load Balancing and Data-Management Services, User's Guide, Sandia National Laboratories Tech. Rep. SAND2007-4748W, 2007.

\bibitem{zoltan-dev}
E. Boman and K. Devine and R. Heaphy and B. Hendrickson and V. Leung and L.A. Riesen and C. Vaughan and U. Catalyurek and D. Bozdag and W. Mitchell, Zoltan v3: Parallel Partitioning, Load Balancing and Data-Management Services, Developer's Guide, Sandia National Laboratories Tech. Rep. SAND2007-4749W, 2007.

\bibitem{nc1}
M. J. Berger, S. H. Bokhari, A partitioning strategy for nonuniform problems
on multiprocessors, IEEE Trans. Computers, 36(5) (1987) 570-580.

\bibitem{nc2}
H. D. Simon, Partitioning of unstructured problems for parallel processing, in:
Proc. Conference on Parallel Methods on Large Scale Structural Analysis and
Physics Applications, Pergammon Press, 1991.

\bibitem{nc3}
V. E. Taylor, B. Nour-Omid, A study of the factorization fill-in for a parallel
implementation of the finite element method, Int. J. Numer. Meth. Engng. 37
(1994) 3809-3823.

\bibitem{nc9}
A. Pothen, H. Simon, K. Liou, Partitioning sparse matrices with eigenvectors
 of graphs, SIAM J. Matrix Anal. 11 (3) (1990) 430-452.


\bibitem{nc10}
T. Bui, C. Jones, A heuristic for reducing fill in sparse matrix factorization,
in: Proc. 6th SIAM Conf. Parallel Processing for Scientific Computing, SIAM,
1993, pp. 445-452.

\bibitem{nc12}
G. Karypis, V. Kumar, A fast and high quality multilevel scheme for
partitioning irregular graphs, Tech. Rep. CORR 95-035, University of
Minnesota, Dept. Computer Science, Minneapolis, MN (June 1995).

\bibitem{nc13}
G. Cybenko, Dynamic load balancing for distributed memory multiprocessors,
J. Parallel Distrib. Comput. 7 (1989) 279-301.

\bibitem{nc14}
Y. Hu, R. Blake, An optimal dynamic load balancing algorithm, Tech. Report
DL-P-95-011, Daresbury Laboratory, Warrington, WA4 4AD, UK (Dec. 1995).

\bibitem{nc15}
E. Leiss, H. Reddy, Distributed load balancing: design and performance
analysis, W.M. Keck Research Computation Laboratory 5 (1989) 205-270.

\bibitem{liuh}
H. Liu, K. Wang, Z. Chen, K. Jordan, J. Luo, H. Deng,
A Parallel Framewrok for Reservoir Simulators on Distributed-memory Supercomputers,
SPE-176045-MS,SPE/IATMI Asia Pacific Oil \& Gas Conference and Exhibition,
Nusa Dua, Indonesia, 20-22 October, 2015.

\bibitem{parmetis}
George Karypis and Vipin Kumar, A Parallel Algorithm for Multilevel Graph Partitioning and Sparse Matrix Ordering,
Journal of Parallel and Distributed Computing, Vol. 48, pp. 71 - 85, 1998

\bibitem{scotch}
F. Pellegrini, static mapping by dual recursive bipartitioning of process and architecture graphs,
Proceedings of SHPCC'94, Knoxville, Tennessee, pages 486-493, IEEE Press, May 1994.

\bibitem{ptscotch}
C. Chevalier, F. Pellegrini, PT-Scotch: A tool for efficient parallel graph ordering,
Parallel Computing, Volume 34, Issues 6¨C8, July 2008, Pages 318¨C331.

\bibitem{chaco}
Bruce Hendrickson, Chaco, Encyclopedia of Parallel Computing, 2011, 248-249, Springer US.

\bibitem{jostle}
Christopher Walshaw and Mark Cross, JOSTLE: multilevel graph partitioning software: an overview,
Mesh partitioning techniques and domain decomposition techniques, Saxe-Coburg Publications, Stirling, Scotland, UK, pp. 27-58, 2007.

\bibitem{Hans}
H. Sagan, Space-Filling Curves. {Springer-Verlag}; 1994.

\bibitem{butz}
A. R. Butz,
Altrnative algorithm for Hilbert's space-filling curve.
{IEEE Transactions on Computers} 1971;
{20}: 424--426.

\bibitem{gold}
L. M. Goldschlager,
Short algorithms for space-filling curves.
{Software---Practice and Experience} 1981;
{11}: 99--100.

\bibitem{witten}
I. H. Witten, B. Wyvill,
On the generation and use of space-filling curves.
{Software---Practice and Experience} 1983;
{13}: 519--525.

\bibitem{cole}
A. J. Cole,
A note on space filling curves.
{Software---Practice and Experience} 1983;
{13}: 1181--1189.

\bibitem{griff}
J. G. Griffiths,
Table-driven algorithms for generating space-filling curves.
{Computer-Aided Design} 1985;
{17}(1): 37--41.

\bibitem{xliu2}
X. Liu, G. F. Schrack,
Encoding and decoding the Hilbert order.
{Software---Practice and Experience} 1996;
{26}(12): 1335--1346.

\bibitem{xliu}
X. Liu, G. F. Schrack, An algorithm for encoding and decoding the 3-D
Hilbert order. {IEEE transactions on image processing} 1997;
{6}: 1333--1337.

\bibitem{fisher}
A. J. Fisher,
A new algorithm for generation hilbert curves.
{Software: Practice and Experience} 1986;
{16}: 5--12.

\bibitem{ningtao}
N. Chen, N. Wang, B. Shi, A new algorithm for encoding and decoding the
Hilbert order. {Software---Practice and Experience} 2007;
{37}(8): 897--908.

\bibitem{kamata}
S. Kamata, R. O. Eason, Y. Bandou, A new algorithm for N-dimensional
Hilbert scanning. {IEEE Trans on Image Processing} 1999;
{8}(7): 964--973.

\bibitem{chenyang}
C. Li, Y. Feng, {Algorithm for analyzing n-dimensional Hilbert
curve} , vol. 3739. {Springer Berlin/Heidelberg}, 2005; 657--662.

\bibitem{PWM}
Peaceman DW, Interpretation of Well-Block Pressures in Numerical Reservoir Simulation,
SPE-6893, 52nd Annual Fall Technical Conference and Exhibition, Denver, 1977.

\bibitem{larry-f}
Ali H. Dogru, Larry Fung, Usuf Middya, Tareq Al-Shaalan, Jorge Alberto Pita,
A Next-Generation Parallel Reservoir Simulator for Giant Reservoirs,
SPE-119272-MS, SPE Reservoir Simulation Symposium, 2-4 February, 2009, The Woodlands, Texas.

\bibitem{amdahl}
David Rodgers, Improvements in multiprocessor system design, ACM SIGARCH Computer Architecture News archive,
New York, NY, USA: ACM. 13 (3): 225?31, 1985.

\bibitem{hsfc-ene}
Nico Reissman, Jan Christian Meyer, Magnus Jahre, A Study of Energy and Locality Effects Using Space-Filling Curves,
Proceedings of the 2014 IEEE International Parallel \& Distributed Processing Symposium Workshops, 815-822,
May 19 - 23, 2014.

\bibitem{hsfc-cluster}
Bongki Moon, H. v. Jagadish, Christos Faloutsos, Joel H. Saltz,
Analysis of the Clustering Properties of the Hilbert Space-Filling Curve,
IEEE Transactions on Knowledge and Data Engineering archive, 13(1), 124-141, 2001.

\bibitem{hsfc-self}
John Hutchinson, Fractals and self-similarity, Indiana University Mathematics Journal, 30, 713-747, 1981.

\bibitem{SPE10}
M. Christie and M. Blun,
Tenth SPE comparative solution project: A comparison of upscaling techniques.
SPE Reservoir Evaluation \& Engineering, 2001, 4(4), 308.

\bibitem{CEL}
Z. Chen, R. E. Ewing, R. D. Lazarov, S. Maliassov, and Y. A. Kuznetsov,
Multilevel Preconditioners for Mixed Methods for Second Order Elliptic Problems, Numerical Linear Algebra with Applications, 3 (5), 427-453, 1996.

\end{thebibliography}
\end{document}